\documentclass[reprint,nofootinbib,amsmath,amssymb,aps,prd,twocolumn]{revtex4-2}
\pdfoutput=1
\usepackage[T1]{fontenc}
\usepackage{amsmath}
\usepackage{siunitx}
\usepackage[colorlinks=true,citecolor=myurlcolor,linkcolor=myurlcolor,urlcolor=myurlcolor]{hyperref}
\usepackage{colortbl}\definecolor{myurlcolor}{rgb}{0,0,0.6}
\usepackage{color} 
\usepackage{graphicx}
\usepackage{dcolumn}
\usepackage{bm}
\usepackage{physics}
\DeclareMathAlphabet{\pazocal}{OMS}{zplm}{m}{n} 
\usepackage{mathrsfs}
\usepackage{dsfont}
\usepackage{mathtools}
\usepackage{comment}

\graphicspath{{./Figures/}}

\newcommand{\zerodel}{.\kern-\nulldelimiterspace}

\renewcommand\bra[1]{{\left<#1\right|}}
\renewcommand\ket[1]{{\left|#1\right>}}
\newcommand{\kket}[1]{\left|\left\zerodel  #1 \right> \right>}

\newcommand{\brakket}[2]{\left\zerodel\left<  #1  \left|   #2  \right> \right>\right\zerodel}
\newcommand{\bbrakket}[2]{\left\zerodel\left< \left<   #1 \left|   #2  \right> \right>\right\zerodel\right\zerodel}

\begin{document}

\preprint{APS/123-QED}

\title{Conditions for Unitarity in Timeless Quantum Theory} 

\author{Simone Rijavec}
\email{simoner@tauex.tau.ac.il}
\affiliation{School of Physics and Astronomy, Tel-Aviv University, Tel-Aviv 69978, Israel}
\affiliation{Clarendon Laboratory, University of Oxford, Parks Road, Oxford OX1 3PU, United Kingdom}

\date{\today}

\begin{abstract}
Quantum timeless approaches solve the problem of time by recovering the usual unitary evolution of quantum theory relative to a clock in a stationary quantum Universe. For some Hamiltonians of the Universe, such as those including an interaction term with the clock, the dynamics is substantially altered and can be non-unitary. This work derives necessary and sufficient conditions for the relative dynamics to be unitary and finds the general form of the unitary evolution operator. A physical interpretation of these conditions is given in terms of the clock's rate. Unitary dynamics is associated with rates that are constant in time and independent of the clock's internal structure.
\end{abstract}

\maketitle

\section{Introduction}
The concept of time as an external parameter in conventional quantum theory clashes with the background-independent character of general relativity and suffers from several explanatory and technical issues, generally termed the ``problem of time'' \cite{kuchar_time_2011,isham_canonical_1993,anderson_problem_2017}. A promising solution to this problem is to remove any appeal to an external time parameter and recover a notion of dynamics \textit{relative} to some systems acting as clocks in an overall stationary quantum Universe \cite{page_evolution_1983}. Crucially, quantum theory interpreted as a \textit{universal} physical theory must already contain an explanation for the relative states underlying such an approach to time \cite{deutsch_three_1986}. Therefore, we can gather the explanations of time that start from a stationary quantum Universe (some of which have been proven to be equivalent \cite{hohn_trinity_2021}) under the term ``timeless quantum theory'' \cite{isham_canonical_1993}.
This approach usually leads to the ordinary dynamics of quantum theory only when the clocks are isolated from the rest of the Universe. The dynamics is substantially altered when the clocks interact with the rest of the Universe \cite{giovannetti_quantum_2015,marletto_evolution_2017,smith_quantizing_2019,smith_quantum_2020,castro-ruiz_quantum_2020,mendes_non-linear_2021,singh_emergence_2025,rijavec_robustness_2023,gemsheim_emergence_2023,favalli_time_2025,kuypers_measuring_2025,cafasso_quantum_2024}, and is even non-unitary for some types of interaction \cite{paiva_flow_2022,paiva_non-inertial_2022,brukner_non-unitarity_nodate}. Non-unitarity was also observed in semiclassical approaches to time \cite{bojowald_effective_2011,bojowald_effective_2011-1,hohn_effective_2012}. The timekeeping of physical clocks is clearly affected if they interact with other systems in specific ways—for instance, when a watch smashes on the ground or melts under intense heat. Such interactions should therefore be prevented.
However, some interactions cannot be shielded \cite{castro_ruiz_entanglement_2017,smith_quantizing_2019,castro-ruiz_quantum_2020} or are required to measure the time of the clock \cite{kuypers_measuring_2025}, while others seem to be inevitable in relativistic settings \cite{smith_quantum_2020}. Therefore, the effect of interactions with the clock in timeless approaches cannot be neglected. Moreover, since unitarity is central to quantum theory, it is crucial to explore the origin of non-unitarity and its role in timeless quantum theory.

In this work, I apply Page and Wootters' approach \cite{page_evolution_1983} to derive necessary and sufficient conditions on the Universe's Hamiltonian for its relative dynamics to be unitary.
These conditions are linked to some physical properties of the clock. Specifically, unitary dynamics is associated with clock rates that are constant in time and independent of the clock's internal structure. 

\section{Review of the Page-Wootters approach}

The starting point is the stationarity condition given by the Wheeler-DeWitt equation \cite{dewitt_quantum_1967}
\begin{equation}
     \hat{H}_U\kket{\Psi_U}=0,
     \label{eq:Wheeler-DeWitt}
 \end{equation}
where $\hat{H}_U$ and $\kket{\Psi_U}$ are the Hamiltonian and the state vector of the Universe, respectively, and the identification of a subsystem $\mathcal{C}$ of the Universe that acts as a clock \cite{page_evolution_1983}. The double-ket notation $\kket{\cdot}$ is a visual aid to denote that the state vector belongs to the Hilbert space of the Universe \cite{giovannetti_quantum_2015}.
In a non-relativistic setting, we can decompose the Hilbert space of the Universe as $\mathscr{H}_U=\mathscr{H}_{\mathcal{C}}\otimes\mathscr{H}_{\mathcal{R}}$, where $\mathscr{H}_{\mathcal{C}}$ and $\mathscr{H}_{\mathcal{R}}$ are the Hilbert spaces of $\mathcal{C}$ and of the rest of the Universe $\mathcal{R}$, respectively, and write $\hat{H}_U$ as
\begin{equation}
     \hat{H}_U=\hat{H}_{\mathcal{C}}\otimes \hat{\mathds{1}}_{\mathcal{R}}+\hat{\mathds{1}}_{\mathcal{C}}\otimes\hat{H}_{\mathcal{R}}+\hat{V},
     \label{eq:H_U}
 \end{equation}
 with $\hat{V}$ an interaction term between  $\mathcal{C}$ and  $\mathcal{R}$.\footnote{In the rest of the paper, the tensor products with the identity operators will often be dropped to improve the readability of the equations. The Hilbert space acted upon by the operators should be clear from the indices and the context.} I will later relax this assumption on the form of $\hat{H}_U$.
In the rest of the discussion, $\mathcal{C}$ is assumed to be an \textit{ideal} clock \cite{hohn_trinity_2021}, meaning that it has a \textit{self-adjoint} operator $\hat{T}_{\mathcal{C}}$ such that $\left[\hat{T}_{\mathcal{C}},\hat{H}_{\mathcal{C}}\right]=i\hbar$, with its eigenstates $\ket{\phi_{\mathcal{C}}(t)}=e^{-i\hat{H}_{\mathcal{C}}(t-t')/\hbar}\ket{\phi_{\mathcal{C}}(t')}$ $\forall t,t' \in \mathds{R}$ corresponding to the different time readings.\footnote{In the rest of the paper $\hbar=1$.} $\mathcal{C}$ is ideal in the sense that the set of states $\left\{\ket{\phi_{\mathcal{C}}(t)}\right\}_{t\in\mathds{R}}$ are orthogonal and so $\mathcal{C}$ can perfectly distinguish any time instant. Such a clock is usually deemed unphysical because $\hat{H}_{\mathcal{C}}$ is unbounded from below. Nevertheless, ideal clocks can be thought of as the limiting case of ever-better realistic clocks. Finally, the states $\left\{\ket{\phi_{\mathcal{C}}(t)}\right\}_{t}$ are not normalizable and thus do not represent physical states. This is analogous to the eigenstates of the position operator in ordinary quantum mechanics and can be mathematically dealt with using the Rigged Hilbert Space formalism \cite{i_m_gelfand_generalized_1964} or introducing a ``physical inner product'' \cite{smith_quantizing_2019}.

Despite the Universe being stationary, we can recover a notion of dynamics by considering the states of the Universe relative to the clock $\mathcal{C}$ showing different times \cite{page_evolution_1983}. When the clock reads the time $t$, the corresponding state of the Universe is given by the following Everettian \cite{everett_relative_1957} relative state
 \begin{equation}
     \kket{\psi_{U}(t)}\coloneqq \hat{\Pi}_t\kket{\Psi_U},
     \label{eq:rel_state}
 \end{equation}
where $\hat{\Pi}_t\coloneqq \ket{\phi_{\mathcal{C}}(t)}\bra{\phi_{\mathcal{C}}(t)}\otimes \hat{\mathds{1}}_{\mathcal{R}} $ is the projector on the state of the clock showing the time $t$.
$\hat{\Pi}_t$ is an ``improper'' projector because $\hat{\Pi}_t^2$ diverges due to the states $\left\{\ket{\phi_{\mathcal{C}}(t)}\right\}_{t}$ being not normalizable.  
Again, this problem can be dealt with using the Rigged Hilbert Space formalism and would not arise for realistic clocks.

If there are no $\mathcal{C}$-$\mathcal{R}$ interactions, that is, $\hat{V}=0$, then Eqs.~\eqref{eq:Wheeler-DeWitt}--\eqref{eq:H_U} imply that 
\begin{equation}
    \kket{\psi_{U}(t)}=e^{-i\hat{H}_{U}t} \kket{\psi_{U}(0)}, \hspace{0.3cm}\forall t \in \mathds{R},
    \label{eq:evo_no_int}
\end{equation} 
recovering the \textit{unitary} evolution of ordinary quantum theory for a system with Hamiltonian $\hat{H}_{U}$ and initial state $\hat{\Pi}_0\kket{\Psi_U}$. In this case, there is the freedom to choose an initial state of the form $\ket{\phi_{\mathcal{C}}(0)}\ket{\psi'_{\mathcal{R}}(0)}$ for any $\ket{\psi'_{\mathcal{R}}(0)}\in \mathscr{H}_{\mathcal{R}}$ \cite{rijavec_robustness_2023}. Note that Eqs.~\eqref{eq:rel_state} and \eqref{eq:evo_no_int} are slightly different from those discussed in the literature, which usually focus on the evolution of $\mathcal{R}$ alone by considering the ``partial'' inner product $\ket{\psi_{\mathcal{R}}(t)}\coloneqq \brakket{\phi_{\mathcal{C}}(t)}{{\Psi_U}}$ \cite{giovannetti_quantum_2015}. Instead, Eqs.~\eqref{eq:rel_state} considers the state of the whole Universe at time $t$, which can also be written as $\kket{\psi_{U}(t)}=\ket{\phi_{\mathcal{C}}(t)} \ket{\psi_{\mathcal{R}}(t)}$. Eq.~\eqref{eq:evo_no_int} implies that $\ket{\psi_{\mathcal{R}}(t)}=e^{-i\hat{H}_{R}t}\ket{\psi_{\mathcal{R}}(0)}$ $\forall t \in \mathds{R}$, so $\ket{\psi_{\mathcal{R}}(t)}$ also evolves unitarily. Unlike $\kket{\psi_{U}(t)}$, $\ket{\psi_{\mathcal{R}}(t)}$ can be chosen to be normalisable and with norm $\norm{\ket{\psi_{\mathcal{R}}(t)}}=1$, $\forall t\in\mathds{R}$ \cite{smith_quantizing_2019}. 

If $\hat{V}\neq0$, $\ket{\psi_{\mathcal{R}}(t)}$ obeys a modified Schr\"odinger equation \cite{smith_quantizing_2019}.
This equation can also be written as
(Appendix \ref{appA:dyn_eq})
\begin{equation}
      \left[i\frac{\text{d}}{\text{d}t}-\hat{H}_{\mathcal{R}} - \hat{V}'\left(t,\frac{\text{d}}{\text{d}t} \right) \right] \ket{\psi_{\mathcal{R}}(t)}=0 ,
    \label{eq:dyn_eq_R}
\end{equation}
where $\hat{V'}$ is an operator-valued function of $t$ and derivatives with respect to $t$.
Similarly, the dynamical equation for $\kket{\psi_{U}(t)}$ is given by
\begin{equation}
      \left[i\frac{\text{d}}{\text{d}t}-\hat{H}_{\mathcal{R}}-\hat{H}_{\mathcal{C}} - \hat{V}''\left(t,\frac{\text{d}}{\text{d}t},\hat{H}_{\mathcal{C}} \right) \right] \kket{\psi_{U}(t)}=0 .
    \label{eq:dyn_eq_U}
\end{equation}
These equations are \textit{linear}, and the maximum order of the time derivatives appearing in them is equal to the maximum power of $\hat{H}_{\mathcal{C}}$ in $\hat{H}_U$.
The solutions of these equations are known in closed form only for some types of interaction \cite{giovannetti_quantum_2015,smith_quantizing_2019,smith_quantum_2020,castro-ruiz_quantum_2020,rijavec_robustness_2023,kuypers_measuring_2025,favalli_time_2025,cafasso_quantum_2024} and can lead to a non-unitary evolution \cite{paiva_flow_2022,paiva_non-inertial_2022}. Note that if $\ket{\psi_{\mathcal{R}}(t)}$ evolves unitarily, then so does $\ket{\Psi_{U}(t)}$, and vice versa (Appendix \ref{appA:unitarity}). Incidentally, this shows that non-unitarity can also appear at the level of the Universe, where no external entangled system has been traced out. Therefore, this effect is different from the non-unitary dynamics of open quantum systems, and is related to the structure of the state of the Universe $\kket{\Psi_U}$ and the way the clock's time states ``slice'' it. 

\section{Sufficient Conditions for Unitarity}
The central result of this work is the following characterization of the types of interaction that lead to a unitary dynamics. First, consider the \textit{self-adjoint} ``rate'' \cite{paiva_non-inertial_2022} or ``redshift'' \cite{castro-ruiz_quantum_2020} operator\footnote{More precisely, one needs to assume that $\hat{H}_{U}$ and $\hat{T}_{\mathcal{C}}$ admit a common invariant dense domain on which the commutator in Eq.~\eqref{eq:alpha_op} is defined, and that $\hat{\alpha}$ admits a \textit{self-adjoint} extension.
Also, note that due to the algebra of $\mathcal{C}$, the commutator $i\left[\cdot,\hat{T}_{\mathcal{C}}\right]$ can be interpreted as a derivative along the diagonal of $\hat{H}_{\mathcal{C}}$ \cite{kuypers_quantum_2022,rijavec_robustness_2023}.}

\begin{equation}
     \hat{\alpha} \coloneqq  i \left[\hat{H}_{U},\hat{T}_{\mathcal{C}}\right].
   \label{eq:alpha_op}
\end{equation}
The following two requirements are sufficient for the dynamics to be unitary:
\begin{gather}
    \left[\hat{T}_{\mathcal{C}},\hat{\alpha}\right]=0,\label{eq:cond1_NEW}\\
    \left[\hat{H}_{U},\hat{\alpha}\right]=0.\label{eq:cond2_NEW}
\end{gather}
Note that these conditions do not rely on the specific form of the Hamiltonian of the Universe of Eq.~\eqref{eq:H_U} and so apply to more general types of constraint (excluding some pathological constraints discussed in Appendix \ref{appA:pathological_cons}). Since we are dealing with unbounded operators, in the rest of the work Eqs.~\eqref{eq:cond1_NEW}-\eqref{eq:cond2_NEW} will be assumed to hold in the strong sense, i.e., that $\hat{\alpha}$ commutes \textit{strongly} with $\hat{T}_{\mathcal{C}}$ and $\hat{H}_{U}$ \cite{reed_methods_2012}. Strong commutativity ensures the existence of a joint spectral measure and hence a joint functional calculus, allowing us to define products of strongly commuting operators unambiguously and control their domains. All the operator equalities are understood on the natural domains defined by the joint functional calculus of the strongly commuting operators involved. Eq.~\eqref{eq:cond1_NEW} in the strong sense implies that $\hat{\alpha}$ must be of the form $\hat{\alpha}=\hat{\alpha}(\hat{T}_{\mathcal{C}})=\int \text{d} t\,\ket{\phi_{\mathcal{C}}(t)}\bra{\phi_{\mathcal{C}}(t)}\otimes \hat{\alpha}_{\mathcal{R}}(t)$, where $t\rightarrow\hat{\alpha}_{\mathcal{R}}(t)$ is a measurable operator-valued function.

Using Eq.~\eqref{eq:cond1_NEW}, $\hat{\Pi}_t\hat{H}_U\kket{\Psi_U}=0$, which follows directly from Eq.~\eqref{eq:Wheeler-DeWitt}, leads to (Appendix \ref{appA:generalized_eq})
\begin{equation}
i\hat{\alpha}\,\frac{\text{d}\kket{\psi_U(t)}}{\text{d}t}= \hat{H}_U\kket{\psi_U(t)},
\label{eq:eq_psiU_1}
\end{equation}
which is a type of generalized Schr\"odinger equation.
There are now two cases. When $\hat{\alpha}$ is invertible, Eq.~\eqref{eq:eq_psiU_1} becomes
\begin{equation}
i\frac{\text{d}\kket{\psi_U(t)}}{\text{d}t}=  \hat{\alpha}^{-1}\hat{H}_U\kket{\psi_U(t)},
\label{eq:eq_psiU_2}
\end{equation}
meaning that $\kket{\psi_U(t)}=\hat{U}(t)\kket{\psi_U(0)}$, with $\hat{U}(t)=e^{-i\hat{\alpha}^{-1}\hat{H}_U t}$. The operator $\hat{\alpha}^{-1}\hat{H}_U$ is defined via the joint functional calculus of the strongly commuting operators $\hat{\alpha}$ and $\hat{H}_U$, which ensures its self-adjointness and thus the \textit{unitarity} of $\hat{U}(t)$.

If $\hat{\alpha}$ is not invertible, then its spectrum must contain 0, that is, $0\in\sigma(\hat{\alpha})$. Calling $\hat{P}^{(0)}=\int \text{d} t\,\ket{\phi_{\mathcal{C}}(t)}\bra{\phi_{\mathcal{C}}(t)}\otimes \hat{P}^{(0)}_{\mathcal{R}}(t)$ the (possibly improper) projector onto the generalized 0-eigenspace of $\hat{\alpha}$ and $\hat{P}^{(+)}=\hat{\mathds{1}}_U-\hat{P}^{(0)}=\int \text{d} t\,\ket{\phi_{\mathcal{C}}(t)}\bra{\phi_{\mathcal{C}}(t)}\otimes \hat{P}^{(+)}_{\mathcal{R}}(t)$ the projector onto its orthogonal complement, so that $\hat{P}^{(0)}+\hat{P}^{(+)}=\hat{\mathds{1}}_U$ and $\hat{P}^{(0)}\hat{P}^{(+)}=\hat{P}^{(+)}\hat{P}^{(0)}=0$, we can multiply Eq.~\eqref{eq:eq_psiU_1} by $\hat{P}^{(0)}$ and $\hat{P}^{(+)}$ to get, respectively,
\begin{gather}
i\hat{\alpha}\,\frac{\text{d}\left(\hat{P}^{(+)}\kket{\psi_U(t)}\right)}{\text{d}t}= \hat{P}^{(+)}\hat{H}_U\kket{\psi_U(t)},
\label{eq:eq_psiU_1A}\\
\hat{P}^{(0)}\hat{H}_U\kket{\psi_U(t)}=0.
\label{eq:eq_psiU_1B}
\end{gather}
The first equation gives the dynamics for the state vector $\hat{P}^{(+)}\kket{\psi_U(t)}=\ket{\phi_{\mathcal{C}}(t)}\hat{P}^{(+)}_{\mathcal{R}}(t)\ket{\psi_{\mathcal{R}}(t)}$; the second imposes a constraint. Note that due to Eq.~\eqref{eq:cond1_NEW} (in the strong sense) the projection preserves the form of the states of the Universe at time $t$ as product states of $\ket{\phi_{\mathcal{C}}(t)}$ and $\ket{\psi_{\mathcal{R}}(t)}$.

In general, the dynamics of $\kket{\psi_U(t)}$ is undetermined.
However, Eq.~\eqref{eq:cond2_NEW} implies that $\hat{P}^{(+)}$ and $\hat{P}^{(0)}$ strongly commute with $\hat{H}_{U}$ so they share a complete set of eigenstates.\footnote{More precisely, their spectral measures commute.} Excluding the pathological cases when $\hat{H}_U$ has 0-eigenstates of the form $\ket{\phi_{\mathcal{C}}(t)}\ket{\chi_{\mathcal{R}}}$ (see Appendix \ref{appA:pathological_cons}), Eq.~\eqref{eq:eq_psiU_1B} implies that $\hat{P}^{(0)}\ket{\Psi_{U}(t)}=\ket{\phi_{\mathcal{C}}(t)}\hat{P}^{(0)}_{\mathcal{R}}(t)\ket{\psi_{\mathcal{R}}(t)}=0$, effectively restricting the allowed states of $\mathcal{R}$ in time. This restriction is well-defined in time because $\hat{P}^{(+)}$ and $\hat{P}^{(0)}$ commute with $\hat{H}_{U}$. Having excluded such states from $\kket{\psi_U(t)}$, Eq.~\eqref{eq:eq_psiU_1A} can be written as
\begin{equation}
i\,\frac{\text{d}\kket{\psi_U(t)}}{\text{d}t}= \hat{\alpha}^{+}\hat{H}_U\kket{\psi_U(t)},
\label{eq:eq_psiU_3}
\end{equation}
where $\hat{\alpha}^{+}$ is the Moore-Penrose inverse of $\hat{\alpha}$ (defined via spectral calculus). The solution of this equation evolves unitarily because $\hat{\alpha}^{+}$ is self-adjoint and if $\hat{\alpha}$ strongly commutes with $\hat{H}_{U}$ then so does $\hat{\alpha}^{+}$ and thus $\hat{\alpha}^{+}\hat{H}_U$ is self-adjoint.

We have thus seen that Eqs.~\eqref{eq:cond1_NEW}--\eqref{eq:cond2_NEW} imply that $\kket{\psi_U(t)}$ evolves with the \textit{unitary} operator
\begin{equation}
    \hat{U}(t)=e^{-i\hat{\alpha}^{+}\hat{H}_U t},
    \label{eq:unitary_op}
\end{equation} 
after possibly restricting $\kket{\psi_{\mathcal{R}}(t)}$ as prescribed by Eq.~\eqref{eq:eq_psiU_1B} if $\hat{\alpha}$ is not invertible ($\hat{\alpha}^{+}=\hat{\alpha}^{-1}$ if $\hat{\alpha}$ is invertible). Consequently, $\ket{\psi_{\mathcal{R}}(t)}$ also evolves unitarily, although its evolution has a more complicated expression; namely,
\begin{equation}
    \ket{\psi_{\mathcal{R}}(t)}=\mathcal{T} \exp\!\left(-i \int_{t_0}^t \text{d}s\, \hat{H}_{\mathcal{R}}^{eff}(s) \right) \ket{\psi_{\mathcal{R}}(t_0)},
\end{equation}
with $\hat{H}_{\mathcal{R}}^{eff}(s)\propto \bra{\phi_{\mathcal{C}}(s)}\hat{\alpha}^{+}\hat{H}_U-\hat{H}_{\mathcal{C}}\ket{\phi_{\mathcal{C}}(s)}$ (see Appendix \ref{appA:dynamics_R}).

$\hat{\alpha}^{+}$ can be seen as modifying the normal rate of the clock, hence the name (inverse) ``rate operator''.
It is also easy to see that $\hat{U}(t)$ leaves $\hat{H}_U$ invariant, $\hat{U}^{\dagger}(t)\hat{H}_U\hat{U}(t)=\hat{H}_U$, meaning that the energy of the Universe is conserved in time, and acts on $\hat{T}_{\mathcal{C}}$ in the following way $\hat{U}^{\dagger}(t)\hat{T}_{\mathcal{C}}\hat{U}(t)=\hat{T}_{\mathcal{C}}+t\cdot\hat{P}^{(+)}$. Lastly, there is the freedom to choose any initial state of the form $\ket{\phi_{\mathcal{C}}(0)}\ket{\chi_{\mathcal{R}}(0)}$ that does not lie in the kernel of $\hat{\alpha}$ (Appendix \ref{appA:freedom}).

\section{Necessity and Constraint Equivalence}
This section shows that while the conditions of Eqs.~\eqref{eq:cond1_NEW} and \eqref{eq:cond2_NEW} are not necessary, they are necessary \textit{up to constraint equivalence}. First, a simple example shows that the conditions are not strictly necessary.
Consider the \textit{interacting} Hamiltonian $\hat{H}'_U=\left(\hat{H}_{\mathcal{C}}^2+\hat{H}_{\mathcal{R}}^2+\delta\right)\cdot\left(\hat{H}_{\mathcal{C}}+\hat{H}_{\mathcal{R}}\right)$, expressed as a product of two commuting factors, with $\delta>0$. 
It is easy to check that $\left[\hat{T}_{\mathcal{C}},\hat{\alpha}\right]\neq 0$ and that $\ket{\psi_{\mathcal{R}}(t)}$ evolves with the unitary operator $e^{-i\hat{H}_{\mathcal{R}}t}$. This means that Eq.~\eqref{eq:cond1_NEW} is not necessary to recover a unitary dynamics. Similar considerations hold for Eq.~\eqref{eq:cond2_NEW}.
Here the reason is easy to understand: the factor $\hat{H}_{\mathcal{C}}^2+\hat{H}_{\mathcal{R}}^2+\delta$ in $\hat{H}'_U$ has no physical relevance for the relative dynamics of the Universe because it is positive definite, so the 0-eigenstates of $\hat{H}'_U$ are all and only the 0-eigenstates of $\hat{H}_{\mathcal{C}}+\hat{H}_{\mathcal{R}}$. Therefore, $\hat{H}'_U$, as a constraint, acts equivalently to $\hat{H}_{\mathcal{C}}+\hat{H}_{\mathcal{R}}$, for which Eqs.~\eqref{eq:cond1_NEW}-\eqref{eq:cond2_NEW} \textit{do} hold.

This is not just a coincidence. It can be proved, more generally, that if $\hat{H}_U$ leads to unitary relative dynamics, then $\hat{H}_U$ is \textit{physically} equivalent to a constraint $\hat{\mathscr{C}}$ for which Eqs.~\eqref{eq:cond1_NEW}--\eqref{eq:cond2_NEW} hold in the \textit{strong} sense. This is proven in Appendix \ref{appA:constraint_equivalence} assuming that $\ket{\Psi_{\mathcal{R}}(t)}$ evolves with a strongly continuous unitary operator $\hat{U}(t,t_0)$ admitting a self-adjoint generator $\hat{X}_{\mathcal{R}}(t)$ which is bounded, $\sup\limits_{t}\norm{\hat{X}_{\mathcal{R}}(t)}<\infty$. Here, physically equivalent means that the \textit{physical} Hilbert space $\mathscr{H}_{phy}(\hat{H}_U)$, i.e. the 0-eigenspace of $\hat{H}_U$, is the same as $\mathscr{H}_{phy}(\hat{\mathscr{C}})$, and consequently $\hat{H}_U$ and $\hat{\mathscr{C}}$ must lead to the same dynamics for $\kket{\psi_U(t)}$ and $\ket{\Psi_{\mathcal{R}}(t)}$.

As a result, if $\hat{H}_U$ is not physically equivalent to a constraint with such properties, then the relative evolution must be non-unitary. For example, the \textit{non-interacting} Hamiltonian of the Universe $\hat{H}''_U=\left(\hat{H}_{\mathcal{C}}+\hat{H}_{\mathcal{R}}\right)\cdot\left(\hat{H}_{\mathcal{C}}-\hat{H}_{\mathcal{R}}\right)$ is not physically equivalent to a constraint satisfying Eqs.~\eqref{eq:cond1_NEW}-\eqref{eq:cond2_NEW} and it indeed leads to a non-unitary relative dynamics. Specifically, $\ket{\psi_{\mathcal{R}}(t)}$ obeys a dynamical equation analogous to the Klein-Gordon equation \cite{diaz_history_2019} and so it evolves, \textit{in general}, non-unitarily. Some \textit{particular} states $\ket{\psi'_{\mathcal{R}}(t)}$ evolve unitarily, but further conditions are required to restrict to only such states \cite{hohn_equivalence_2021}.

\section{Physical Interpretation}
What is the physical interpretation of Eqs.~\eqref{eq:cond1_NEW} and \eqref{eq:cond2_NEW}? To understand this, consider the meaning of $\hat{\alpha}$. This operator was defined in \cite{paiva_non-inertial_2022} as the ``time rate of the clock'' $\mathcal{C}$ from the perspective of another non-interacting clock $\mathcal{C}_2$. As $\hat{\alpha}$ is not a c-number, it should be more aptly called a rate \textit{operator}. To see how it is connected to the rate of $\mathcal{C}$ from the perspective of $\mathcal{C}_2$, assume that $\mathcal{C}_2$ is an ideal clock and consider the Hamiltonian of the Universe  $\hat{H}_U=\hat{H}_{\mathcal{C}} + \hat{H}_{\mathcal{R}}+\hat{V}_{\mathcal{C}\mathcal{R}} + \hat{H}_{\mathcal{C}_2}$, with $\hat{V}_{\mathcal{C}\mathcal{R}}$ an interaction term between $\mathcal{C}$ and $\mathcal{R}$ only. Similarly to Eqs.~\eqref{eq:rel_state}--\eqref{eq:evo_no_int}, the state of $\mathcal{CR}$ when $\mathcal{C}_2$ shows the time $t$ is given by $\ket{\psi_{\mathcal{CR}}(t)}=e^{-i\hat{H}_{\mathcal{CR}}t}\ket{\psi_{\mathcal{CR}}(0)}$, with $\hat{H}_{\mathcal{CR}}\coloneqq \hat{H}_{\mathcal{C}} + \hat{H}_{\mathcal{R}}+\hat{V}_{\mathcal{C}\mathcal{R}}$. Due to $\mathcal{C}$'s interaction with $\mathcal{R}$, $\ket{\psi_{\mathcal{CR}}(t)}$ will be, in general, an entangled state consisting of a superposition of states with different time readings of $\mathcal{C}$, $\ket{\psi_{\mathcal{CR}}(t)}=\int_{-\infty}^{+\infty}\text{d}\tau\,\ket{\phi_{\mathcal{C}}(\tau)}\braket{\phi_{\mathcal{C}}(\tau)}{\psi_{\mathcal{CR}}(t)}$. The average time reading of $\mathcal{C}$ when $\mathcal{C}_2$ reads $t$ is given by
\begin{equation}
     \overline{\tau_{\mathcal{C}}}(t)\coloneqq\frac{\bra{\psi_{\mathcal{CR}}(t)}\hat{T}_{\mathcal{C}}\ket{\psi_{\mathcal{CR}}(t)}}{\braket{\psi_{\mathcal{CR}}(t)}{\psi_{\mathcal{CR}}(t)}},
\end{equation}
so the rate of $\mathcal{C}$ from the perspective of $\mathcal{C}_2$ is (Appendix \ref{appA:clock})
\begin{equation}
\alpha(t) \coloneqq  \frac{\text{d}\overline{\tau_{\mathcal{C}}}(t)}{\text{d}t}=\frac{\bra{\psi_{\mathcal{CR}}(t)} \hat{\alpha} \ket{\psi_{\mathcal{CR}}(t)}}{\braket{\psi_{\mathcal{CR}}(t)}{\psi_{\mathcal{CR}}(t)}}.
    \label{eq4:gamma_c}
\end{equation}
This equation connects the rate operator $\hat{\alpha}$ of Eq.~\eqref{eq:alpha_op} to the real-valued, time-dependent rate $\alpha(t)$ (in special relativity we would call this quantity $1/\gamma(t)$).\footnote{Both the numerator and the denominator in these equations are divergent due to the states of the clock $\mathcal{C}$ being non-normalizable. Here, I assume that these divergencies cancel out, as they would if $\mathcal{C}$ were a realistic clock.}

Now, Eq.~\eqref{eq:cond2_NEW} implies that (Appendix \ref{appA:clock})
\begin{equation}
     \frac{\text{d}\alpha(t)}{\text{d}t} = 0, \hspace{0.3cm} \forall t\in\mathds{R},
     \label{eq:rate_der}
\end{equation}
meaning that the rate of the clock is constant in time, $\alpha(t)=\alpha(0)$ for all $t\in\mathds{R}$. The converse is also true: if $\alpha(t)$ is constant in time for any choice of initial state $\ket{\psi_{\mathcal{CR}}(0)}$ (this is allowed because $\mathcal{CR}$ does not interact with $\mathcal{C}_2$), then $\left[\hat{H}_{U},\hat{\alpha}\right]=0$. Eq.~\eqref{eq:rate_der} is also equivalent to
\begin{equation}
     \overline{\tau_{\mathcal{C}}}(t)=\overline{\tau_{\mathcal{C}}}(0)+\alpha(0) t,
     \label{eq:time_transformation}
\end{equation}
showing that the times of $\mathcal{C}$ and $\mathcal{C}_2$ are related by a linear transformation.
The same would hold for any other clock not interacting with $\mathcal{C}$. 

Eq.~\eqref{eq:cond2_NEW} also implies that the variance of the time readings of $\mathcal{C}$ from the perspective of $\mathcal{C}_2$ at time $t$ is given by 
\begin{equation}
     \sigma^2_{\tau_{\mathcal{C}}}(t)= \sigma^2_{\tau_{\mathcal{C}}}(0)+t^2 \sigma^2_{\alpha}(0)+2t\,\text{Cov}(\alpha(0),\tau_{\mathcal{C}}(0)) ,
     \label{eq:variance_clock}
\end{equation}
with $\sigma^2_{\tau_{\mathcal{C}}}(0)$, $\sigma^2_{\alpha}(0)$ and $\text{Cov}(\alpha(0),\tau_{\mathcal{C}}(0))$ the variances and covariance of $\tau_{\mathcal{C}}$ and $\alpha$ at time $0$, respectively (Appendix \ref{appA:clock}).
When $\ket{\psi_{\mathcal{CR}}(0)}$ is an eigenstate of $\hat{T}_{\mathcal{C}}$, this reduces to $\sigma^2_{\tau_{\mathcal{C}}}(t)=t^2 \sigma^2_{\alpha}(0)$ and the relative uncertainty on the readings of $\mathcal{C}$ from the perspective of $\mathcal{C}_2$ is  $\sigma_{\tau_{\mathcal{C}}}/t=\sigma_{\alpha}$, which is constant in time.

What about Eq.~\eqref{eq:cond1_NEW}? This condition can be interpreted as imposing that the rate operator $\hat{\alpha}$ must be independent of the internal structure of the clock. This is because the equation (in the strong sense) implies that $\hat{\alpha}$ is an operator-valued function of $\hat{T}_{\mathcal{C}}$ only and thus does not contain $\hat{H}_{\mathcal{C}}$, which describes the internal structure of the clock. This is similar to relativity, where time dilation effects are independent of the clocks chosen to measure the time.\footnote{We can imagine a Hamiltonian of the Universe $\hat{H}_U=\gamma^{-1}\hat{H}_{\mathcal{C}}+\hat{H}_{\mathcal{R}}$ with $\gamma^{-1}\in\mathds{R}$, for which $\hat{\alpha}=\gamma^{-1}\hat{\mathds{1}}_U$. However, in this case we would have to rescale the time states $\ket{\phi_{\mathcal{C}}(t)}$ with the same factor to preserve their interpretation. This would change $\hat{T}_{\mathcal{C}}$ to $\gamma\hat{T}_{\mathcal{C}}$ and restore the ``natural'' rate of the clock $\hat{\alpha}=\hat{\mathds{1}}_U$.}

\section{Conclusions}
We have seen that the following statements are equivalent:
\begin{enumerate}
    \item $\kket{\psi_{U}(t)}$ and $ \ket{\psi_{\mathcal{R}}(t)}$ evolve unitarily.
    \item $\hat{H}_U$ is physically equivalent to a constraint for which Eqs.~\eqref{eq:cond1_NEW} and \eqref{eq:cond2_NEW} hold (excluding the pathological constraints discussed in Appendix \ref{appA:pathological_cons}).
    \item The rate of the clock $\mathcal{C}$ is constant with respect to any other non-interacting clock and does not depend on the internal structure of $\mathcal{C}$.
\end{enumerate}
This characterizes the types of Hamiltonians of the Universe that lead to a unitary dynamics from both the mathematical and the interpretational point of view.
This characterization relies on the clock $\mathcal{C}$ being an ideal clock, so it should also hold in the limiting case of ever more precise realistic clocks.  Therefore, we can expect its insights on unitarity in timeless quantum theory to be general. Nevertheless, it would be useful to consider how these results extend to non-ideal clocks, which are often treated using POVMs rather than projective measurements \cite{busch_energy-time_1990,smith_quantum_2020,hohn_trinity_2021,chataignier_relational_2026}, and where further breakdowns of unitarity might appear \cite{hausmann_measurement_2025}. Closely related questions arise in the framework of quantum reference frames, where time is treated relationally \cite{giacomini_quantum_2019,hohn_how_2020,vanrietvelde_change_2020}. I leave these investigations for future work.

What to make of non-unitarity? The two conditions found in this work point to two different origins of this effect. Hamiltonians of the Universe that violate Eq.~\eqref{eq:cond1_NEW} make the rate of the clock depend on its internal structure, meaning that the flow of time changes with the choice of clock, and so the effect is not general but clock-specific. This might happen for some types of interactions with the clock, but we have also observed this for a non-interacting relativistic Hamiltonian of the Universe. In this latter case, non-unitarity is due to the relativistic quantum mechanical treatment, similarly to the non-unitarity of the Klein-Gordon equation, and the problem might be solved by adopting a quantum field theoretic timeless approach (see \cite{hohn_matter_2024} for a first step in this direction).

Hamiltonians that satisfy Eq.~\eqref{eq:cond1_NEW} but violate Eq.~\eqref{eq:cond2_NEW} seem to lead to a different type of non-unitarity. Specifically, they are related to clock rates that change in time but do so independently of the clock choice. In this case, the origin of non-unitarity can be intuitively grasped in the following way. Assume for simplicity that $\hat{\alpha}$ is invertible. The evolution operator of $\kket{\psi_U(t)}$ is still given by $\hat{U}(t)=e^{-i\hat{\alpha}^{-1}\hat{H}_U t}$, but with $\hat{U}(t)$ \textit{non-unitary} because $\hat{\alpha}^{-1}$ and $\hat{H}_U$ do not commute. $\hat{\alpha}^{-1}$ can be seen as changing the evolution rate of the different eigenstates of $\hat{\alpha}^{-1}$ appearing in $\kket{\psi_U(t)}$. If $\hat{\alpha}^{-1}$ and $\hat{H}_U$ commuted, then they would share a common set of eigenstates $\left\{\kket{\chi_j}\right\}_j$ (assuming the set is discrete for ease of notation) such that $\hat{\alpha}^{-1}\kket{\chi_j}=\alpha^{-1}_j\kket{\chi_j}$ and $\hat{H}_U\kket{\chi_j}=E_j\kket{\chi_j}$, and thus $\kket{\psi_U(t)}$ could be expressed 
as $\kket{\psi_U(t)}=\sum_j\lambda_j e^{-i \alpha^{-1}_j E_j t}\kket{\chi_j}$, showing that $\hat{\alpha}^{-1}$ affects the evolution rate of the different eigenstates in a way that is non-trivial but preserves unitarity. However, if $\hat{\alpha}^{-1}$ and $\hat{H}_U$ do not commute, then $\hat{H}_U$ mixes the different eigenstates of $\hat{\alpha}^{-1}$, so a state that starts evolving with a certain rate ends up in a superposition of states evolving with different rates. The result is a non-unitary evolution. 

Under the condition of Eq.~\eqref{eq:cond1_NEW}, preventing the clock from coupling to other systems would ensure unitarity because we would have $\hat{\alpha}=\hat{\mathds{1}}_U$. If the clock were a composite system, it would be sufficient to prevent the coupling between the degrees of freedom responsible for the clock's timekeeping and the rest of the Universe. However, some interactions with the clock's internal degrees of freedom may be inevitable \cite{castro_ruiz_entanglement_2017} while others lead to desirable relativistic time dilation effects \cite{smith_quantizing_2019}. A straightforward generalization of these types of interactions leads to non-unitarity. Consider, for instance, $\hat{H}_U=\hat{H}^{int}_{\mathcal{C}}+\frac{\left(\hat{P}^{CM}_{\mathcal{C}}\right)^2}{2 m_{\mathcal{C}}}+\hat{H}_{\mathcal{R}} + \Lambda(\hat{X}^{CM}_{\mathcal{C}})\hat{H}^{int}_{\mathcal{C}}\hat{H}_{\mathcal{R}}$, with $\hat{H}^{int}_{\mathcal{C}}$ the Hamiltonian of the internal degrees of freedom of the clock, $\hat{P}^{CM}_{\mathcal{C}}$ and $\hat{X}^{CM}_{\mathcal{C}}$ the momentum and position of its centre of mass, and the interaction term obtained using the mass-energy equivalence principle 
 in a Newtonian-like potential between $\mathcal{C}$ and $\mathcal{R}$ \cite{castro_ruiz_entanglement_2017} which is function of $\hat{X}^{CM}_{\mathcal{C}}$. The rate operator $\hat{\alpha}=\hat{\mathds{1}}_U+\Lambda(\hat{X}^{CM}_{\mathcal{C}})\hat{H}_{\mathcal{R}}$ does not commute with $\hat{H}_U$ and so the evolution is non-unitary despite $\hat{H}_U$ describing a simple and natural physical scenario of a clock moving under the influence of a Newtonian-like potential.
 
Overall, non-unitarity appears to be a genuine feature of timeless quantum theory related to well-defined physical properties of the clock. It is now crucial to understand how this effect can be reconciled with the physical interpretation of unitarity in standard quantum theory.


\subparagraph{Acknowledgments.}
The author is grateful to an anonymous referee for a careful and constructive review, which led to substantial improvements in the mathematical foundations of the manuscript.
This work has been supported in part by the Israel Science Foundation Grant No. 1713/24.

\bibliography{main}

\clearpage
\onecolumngrid
\appendix 
\section{General dynamical equation}\label{appA:dyn_eq}
Since $\hat{H}_{\mathcal{C}}$ and $\hat{T}_{\mathcal{C}}$ generate the algebra of observables of ${\mathcal{C}}$, the interaction term of the Hamiltonian of the Universe $\hat{V}$ in Eq.~\eqref{eq:H_U} can be written as a sum of products of $\hat{H}_{\mathcal{C}}$, $\hat{T}_{\mathcal{C}}$, and some observables of $\mathcal{R}$. Let me denote this as $\hat{V}=\hat{V}(\hat{T}_{\mathcal{C}},\hat{H}_{\mathcal{C}})$. We can then rewrite $\hat{V}(\hat{T}_{\mathcal{C}},\hat{H}_{\mathcal{C}})$ as $\hat{V}'(\hat{T}_{\mathcal{C}},\hat{H}_{\mathcal{C}})$ where all $\hat{T}_{\mathcal{C}}$ are moved to the left of $\hat{H}_{\mathcal{C}}$ using the fact that $\left[\hat{T}_{\mathcal{C}},\hat{H}_{\mathcal{C}}\right]=i\hbar$. Now, $\hat{H}_{\mathcal{C}}$ acts on the time states of the clock as $\hat{H}_{\mathcal{C}}\ket{\phi_{\mathcal{C}}(t)}=i\frac{\text{d}\ket{\phi_{\mathcal{C}}(t)}}{\text{d}t}$, so when taking the partial inner product $\bra{\phi_{\mathcal{C}}(t)}\hat{H}_U\kket{\Psi_U}=0$, with $\hat{H}_U$ expressed using  $\hat{V}'(\hat{T}_{\mathcal{C}},\hat{H}_{\mathcal{C}})$, we can simply substitute $\hat{T}_{\mathcal{C}}$ with $t$ and $\hat{H}_{\mathcal{C}}$ with $i\frac{\text{d}}{\text{d}t}$ to get Eq.~\eqref{eq:dyn_eq_R}.
Similarly, using $\hat{\Pi}_t\hat{H}_U\kket{\Psi_U}=0$ and the fact that $i\frac{\text{d}\hat{\Pi}_t}{\text{d}t}=\left[\hat{H}_{\mathcal{C}},\hat{\Pi}_t\right]$, we get Eq.~\eqref{eq:dyn_eq_U}.

Consider, for instance, $\hat{V}=\hat{T}_{\mathcal{C}}\hat{H}_{\mathcal{C}}\hat{T}_{\mathcal{C}}\otimes \hat{O}_{\mathcal{R}}$.
We can rewrite it as $\hat{V}'(\hat{T}_{\mathcal{C}},\hat{H}_{\mathcal{C}})=\hat{T}^2_{\mathcal{C}}\hat{H}_{\mathcal{C}}\otimes \hat{O}_{\mathcal{R}}-i\hat{T}_{\mathcal{C}}\otimes \hat{O}_{\mathcal{R}}$ and so we have
\begin{equation}
   \left[i\frac{\text{d}}{\text{d}t}-\hat{H}_{\mathcal{R}} + i t^2\hat{O}_{\mathcal{R}}\frac{\text{d}}{\text{d}t} + it \hat{O}_{\mathcal{R}}\right] \ket{\psi_{\mathcal{R}}(t)}=0,
\end{equation}
and
\begin{equation}
   \left[i\frac{\text{d}}{\text{d}t}-\hat{H}_{\mathcal{R}} -\hat{H}_{\mathcal{C}} + i t^2\hat{O}_{\mathcal{R}}\frac{\text{d}}{\text{d}t}  
  - t^2\hat{H}_{\mathcal{C}}\hat{O}_{\mathcal{R}}+ it \hat{O}_{\mathcal{R}}\right] \ket{\psi_{U}(t)}=0.
\end{equation}
If the $n$-th power of $\hat{H}_{\mathcal{C}}$ appeared in $\hat{V}$, then these equations would contain time derivatives up to order $n$.

\section{Global and local unitarity}\label{appA:unitarity}
As discussed in the body of this work, we need to be careful when talking about unitarity in timeless quantum theory because the constraint of Eq.~\eqref{eq:Wheeler-DeWitt} can exclude some time states of $\mathcal{R}$/$U$ and so the evolution operator of $\ket{\psi_{\mathcal{R}}(t)}$ or $\ket{\psi_{U}(t)}$ does not necessarily need to be unitary but just needs to act unitarily on the \textit{subspace} of allowed time states of $\mathcal{R}$/$U$. Therefore, there will be an equivalence class of operators that act unitarily on such subspace and, in the following, \textit{unitary evolution} will refer to the fact that the evolution operator of $\ket{\psi_{\mathcal{R}}(t)}$ or $\ket{\psi_{U}(t)}$ belongs to such an equivalence class.

With this clarification, we can easily see that if $\ket{\psi_{\mathcal{R}}(t)}$ evolves unitarily, then so does $\kket{\psi_U(t)}$. This is because if $\ket{\psi_{\mathcal{R}}(t)}=\hat{V}(t,t_0)\ket{\psi_{\mathcal{R}}(t_0)}$ with $\hat{V}(t,t_0)$ unitary, then $\kket{\psi_{U}(t)}=\hat{W}(t,t_0)\kket{\psi_{U}(t_0)}$ with $\hat{W}(t,t_0)\coloneqq e^{-i\hat{H}_{\mathcal{C}}(t-t_0)}\otimes \hat{V}(t,t_0)$ which is also unitary.  
The converse is also true. To see this, assume that $\ket{\psi_{\mathcal{R}}(t)}$ evolves non-unitarily, meaning that $\ket{\psi_{\mathcal{R}}(t)}=\hat{V'}(t)\ket{\psi_{\mathcal{R}}(0)}$ with $\hat{V'}(t)$ not equivalent to an operator acting unitarily on the subspace of allowed states of $\mathcal{R}$. The states of the Universe at different times can then be expressed as $\kket{\psi_{U}(t)}=\hat{W'}(t)\kket{\psi_{U}(0)}$ with $\hat{W'}(t)\coloneqq e^{-i\hat{H}_{\mathcal{C}}t}\otimes \hat{V'}(t)$ which is also not equivalent to an operator acting unitarily on the subspace of the allowed time states of $U$.

\section{Generalized Schr\"odinger equation}\label{appA:generalized_eq}
The first step to derive Eq.~\eqref{eq:eq_psiU_1} is to write the improper projector $\ket{\phi_{\mathcal{C}}(t)}\bra{\phi_{\mathcal{C}}(t)}$ in the following way
\begin{equation}
    \ket{\phi_{\mathcal{C}}(t)}\bra{\phi_{\mathcal{C}}(t)}=\int_{-\infty}^{+\infty}\frac{\text{d}\varepsilon}{2\pi} \,e^{i\varepsilon(\hat{T}_{\mathcal{C}}-t)}=\delta(\hat{T}_{\mathcal{C}}-t).
    \label{eq:projector_t}
\end{equation}
One can easily check that $\int_{-\infty}^{+\infty}\frac{\text{d}\varepsilon }{2\pi}\,e^{i\varepsilon(\hat{T}_{\mathcal{C}}-t)}\ket{\phi_{\mathcal{C}}(t')}=\int_{-\infty}^{+\infty}\frac{\text{d}\varepsilon}{2\pi} \,e^{i\varepsilon(t'-t)}\ket{\phi_{\mathcal{C}}(t')}=\delta(t'-t)\ket{\phi_{\mathcal{C}}(t')}$, as expected. Therefore $\hat{\Pi}_t=\int_{-\infty}^{+\infty}\frac{\text{d}\varepsilon }{2\pi}\,e^{i\varepsilon(\hat{T}_{\mathcal{C}}-t)}\otimes\hat{\mathds{1}}_{\mathcal{R}}$.
We can use this expression and Eq.~\eqref{eq:Wheeler-DeWitt} to get (omitting $\hat{\mathds{1}}_{\mathcal{R}}$ for ease of notation)
\begin{equation}
    0=\hat{\Pi}_t\hat{H}_U\kket{\Psi_U}=\int_{-\infty}^{+\infty}\frac{\text{d}\varepsilon }{2\pi}\,e^{i\varepsilon(\hat{T}_{\mathcal{C}}-t)}\hat{H}_U\kket{\Psi_U}
    =\int_{-\infty}^{+\infty}\frac{\text{d}\varepsilon}{2\pi} \,e^{i\varepsilon(\hat{T}_{\mathcal{C}}-t)}\hat{H}_U e^{-i\varepsilon(\hat{T}_{\mathcal{C}}-t)}e^{i\varepsilon(\hat{T}_{\mathcal{C}}-t)}\kket{\Psi_U}.
    \label{eq:calc1}
\end{equation}
But we also have 
\begin{equation}
e^{i\varepsilon(\hat{T}_{\mathcal{C}}-t)}\hat{H}_U e^{-i\varepsilon(\hat{T}_{\mathcal{C}}-t)}= \hat{H}_U+i\varepsilon\left[\hat{T}_{\mathcal{C}},\hat{H}_U\right]+\frac{\left(i\varepsilon\right)^2}{2!}\left[\hat{T}_{\mathcal{C}},\left[\hat{T}_{\mathcal{C}},\hat{H}_U\right]\right]+\dots = \hat{H}_U+i\varepsilon\left[\hat{T}_{\mathcal{C}},\hat{H}_U\right]=\hat{H}_U-\varepsilon\hat{\alpha},
   \label{eq:commutator1}
\end{equation}
where the second equality follows from Eqs.~\eqref{eq:alpha_op} and \eqref{eq:cond1_NEW}.
Therefore, Eq.~\eqref{eq:calc1} becomes
\begin{equation}
    0=\int_{-\infty}^{+\infty}\frac{\text{d}\varepsilon}{2\pi} \,\left(\hat{H}_U-\varepsilon\hat{\alpha}\right)e^{i\varepsilon(\hat{T}_{\mathcal{C}}-t)}\kket{\Psi_U}
    =\left(\hat{H}_U-i\hat{\alpha}\frac{\text{d}}{\text{d}t}\right)\int_{-\infty}^{+\infty}\frac{\text{d}\varepsilon}{2\pi} \,e^{i\varepsilon(\hat{T}_{\mathcal{C}}-t)}\kket{\Psi_U}
    =\left(\hat{H}_U-i\hat{\alpha}\frac{\text{d}}{\text{d}t}\right)\kket{\psi_U(t)}.
\end{equation}
Rearranging the terms we get Eq.~\eqref{eq:eq_psiU_1}.

\section{Pathological constraints}\label{appA:pathological_cons}
The timeless approach considered in this work is ill-defined for some pathological types of constraints.
Such constraints do not lead to a good notion of dynamics, so the results of this work do not apply to them.
In general, any constraint $\hat{\mathscr{C}}$ that has a 0-eigenstate of the form $\ket{\phi_{\mathcal{C}}(t)}\ket{\chi_{\mathcal{R}}}$ for some $t$ and $\ket{\chi_{\mathcal{R}}}\in\mathscr{H}_{\mathcal{R}}$ is problematic. This is because $\kket{\Psi_U}=\ket{\phi_{\mathcal{C}}(t)}\ket{\chi_{\mathcal{R}}}$ is an allowed state of the Universe but does not lead to any dynamics and, in fact, allows only one time instant to exist.
Moreover, if there is more than one such eigenstate for different values of $t$, then we can consider the state of the Universe $\kket{\Psi_U}=\sum_j \lambda(t_j) \ket{\phi_{\mathcal{C}}(t_j)}\ket{\chi(t_j)}$. Since the $\lambda(t_j)$ can be chosen arbitrarily, the norm of $\ket{\psi_{\mathcal{R}}(t)}$ can be changed arbitrarily and so the operator connecting the different $\ket{\psi_{\mathcal{R}}(t_j)}$ (if it exists) is, in general, not unitary.

An example of such pathological constraints is the class of constraints $\hat{\mathscr{C}}$ that do not contain $\hat{H}_{\mathcal{C}}$ but only $\hat{T}_{\mathcal{C}}$.
The 0-eigenstates of $\hat{\mathscr{C}}$  must all be of the form $\ket{\phi_{\mathcal{C}}(t)}\ket{\chi(t)}$ because $\hat{\mathscr{C}}$ commutes with $\hat{T}_{\mathcal{C}}$ (which is non-degenerate). 
The states $\kket{\psi_U(t)}$ deriving from such constraints do not obey a dynamical equation as in Eq.~\eqref{eq:dyn_eq_U} because there is no derivative with respect to time. This is reflected in the fact that $\hat{\alpha}=0$. 

Another problematic class of constraints consists of those that can be \textit{factorized} as
$\hat{\mathscr{C}}=\hat{\mathscr{C}}_1\cdot \hat{\mathscr{C}}_2$ with $\hat{\mathscr{C}}_2$ of the form $\hat{\mathds{1}}_{\mathcal{C}}\otimes\hat{\mathscr{C}}_2$ having at least one 0-eigenstate ($\hat{\mathscr{C}}_1$ and $\hat{\mathscr{C}}_2$ must commute because $\hat{\mathscr{C}}$ is self-adjoint). In this case, there is a set of 0-eigenstates of $\hat{\mathscr{C}}$ of the form $\left\{\ket{\varphi}\ket{\chi_{j}}\right\}_j$, with $\hat{\mathscr{C}}_2\ket{\chi_{j}}=0$ $\forall j$, for \textit{any} $\ket{\varphi}\in\mathscr{H}_{\mathcal{C}}$, so, in particular, any $\ket{\phi_{\mathcal{C}}(t)}\ket{\chi_{j}}$ will be a 0-eigenstate of $\hat{\mathscr{C}}$ for any $t\in\mathds{R}$.

\section{Reduced relative dynamics}\label{appA:dynamics_R}
Let us start with Eq.~\eqref{eq:eq_psiU_3} which we can write as
\begin{equation}
i\ket{\phi_{\mathcal{C}}(t)}\frac{\text{d}\ket{\psi_{\mathcal{R}}(t)}}{\text{d}t}= \left(\hat{\alpha}^{+}\hat{H}_U - \hat{H}_{\mathcal{C}}\right) \ket{\phi_{\mathcal{C}}(t)}\ket{\psi_{\mathcal{R}}(t)}.
\end{equation}
If we multiply this equation on the left with $\hat{\Pi}_{t'}$ and integrate in $t'$ we get
\begin{equation}
i\,\frac{\text{d}\ket{\psi_{\mathcal{R}}(t')}}{\text{d}t'}=\int \text{d}t'\, \bra{\phi_{\mathcal{C}}(t')}\hat{\alpha}^{+}\hat{H}_U - \hat{H}_{\mathcal{C}}\ket{\phi_{\mathcal{C}}(t)}\ket{\psi_{\mathcal{R}}(t)},
\end{equation}
which is similar to the equation appearing in \cite{smith_quantizing_2019}.
However, in this case Eqs.~\eqref{eq:alpha_op}-\eqref{eq:cond2_NEW} imply that
\begin{equation}
    \left[\hat{\alpha}^{+}\hat{H}_U - \hat{H}_{\mathcal{C}},\hat{T}_{\mathcal{C}}\right]=-i\hat{\alpha}^{+}\hat{\alpha}+i=i(\hat{P}^{(+)}-1)=i\hat{P}^{(0)},
\end{equation}
from which follows that $(t-t')\bra{\phi_{\mathcal{C}}(t')}\hat{\alpha}^{+}\hat{H}_U - \hat{H}_{\mathcal{C}}\ket{\phi_{\mathcal{C}}(t)}\ket{\psi_{\mathcal{R}}(t)}=i(t-t')\hat{P}^{(0)}(t)\ket{\psi_{\mathcal{R}}(t)}=0$, $\forall t,t'$, because $\ket{\psi_{\mathcal{R}}(t)}$ is restricted to the subspace orthogonal to $\hat{P}^{(0)}$. Therefore, after renaming the variable $t'$, 
\begin{equation}
i\,\frac{\text{d}\ket{\psi_{\mathcal{R}}(t)}}{\text{d}t}=\hat{H}_{\mathcal{R}}^{eff}(t)\ket{\psi_{\mathcal{R}}(t)},
\end{equation}
with $\hat{H}_{\mathcal{R}}^{eff}(t) \propto \bra{\phi_{\mathcal{C}}(t)}\hat{\alpha}^{+}\hat{H}_U - \hat{H}_{\mathcal{C}}\ket{\phi_{\mathcal{C}}(t)}$. The solution to this equation is 
\begin{equation}
    \ket{\psi_{\mathcal{R}}(t)}=\mathcal{T} \exp\!\left(-i \int_{t_0}^t \text{d}s\, \hat{H}_{\mathcal{R}}^{eff}(s) \right) \ket{\psi_{\mathcal{R}}(t_0)}.
\end{equation}
Also note that we can write $\hat{H}_{\mathcal{R}}^{eff}(t)$, with an abuse of notation, as $\hat{H}_{\mathcal{R}}^{eff}(t)= \bra{\phi_{\mathcal{C}}(t)}\hat{\alpha}^{+}\hat{H}_U - \hat{H}_{\mathcal{C}}\ket{\phi_{\mathcal{C}}(t)}/\delta(0)$.

\section{Freedom in the choice of initial state}\label{appA:freedom}
It is possible to construct a stationary state of the Universe starting from any state $\kket{\Theta}\in \mathscr{H}_U$ in the following way
\begin{equation}
    \kket{\Psi_U}=\int_{-\infty}^{+\infty}\text{d}t\,e^{-i\hat{H}_U t}\kket{\Theta}.
\end{equation}
In particular,
\begin{equation}
    \kket{\Psi_U}=\int_{-\infty}^{+\infty}\text{d}t\,e^{-i\hat{H}_U t}\ket{\phi_{\mathcal{C}}(0)}\ket{\chi_{\mathcal{R}}},
    \label{eq:construction_Psi}
\end{equation}
is also a stationary state of the Universe for any $\ket{\chi_{\mathcal{R}}}\in\mathscr{H}_{\mathcal{R}}$.
While $\kket{\Psi_U}$ in Eq.~\eqref{eq:construction_Psi} is constructed from  $\ket{\phi_{\mathcal{C}}(0)}\ket{\chi_{\mathcal{R}}}$, it is not necessarily true that the state of the Universe at time $t=0$ is $\ket{\phi_{\mathcal{C}}(0)}\ket{\chi_{\mathcal{R}}}$, that is, we could have $\hat{\Pi}_0\kket{\Psi_U}\neq\ket{\phi_{\mathcal{C}}(0)}\ket{\chi_{\mathcal{R}}}$. However, if $\hat{H}_U$ satisfies Eqs.~\eqref{eq:cond1_NEW}--\eqref{eq:cond2_NEW} then we can choose the state used to construct $\kket{\Psi_U}$ so that $\hat{\Pi}_0\kket{\Psi_U}$ is equal to $\ket{\phi_{\mathcal{C}}(0)}\ket{\chi_{\mathcal{R}}}$ for any $\ket{\chi_{\mathcal{R}}}\in\mathscr{H}_{\mathcal{R}}$ as long as $\ket{\phi_{\mathcal{C}}(0)}\ket{\chi_{\mathcal{R}}}$ does not lie in the kernel of $\hat{\alpha}$.
To see this, consider
\begin{equation}
   \hat{\Pi}_0 \kket{\Psi_U}=\int_{-\infty}^{+\infty}\text{d}t\,\hat{\Pi}_0 e^{-i\hat{H}_U t}\ket{\phi_{\mathcal{C}}(0)}\ket{\chi_{\mathcal{R}}}
\end{equation}
Using Eq.~\eqref{eq:projector_t}, we can write this as
\begin{multline}
\hat{\Pi}_0 \kket{\Psi_U}=\int_{-\infty}^{+\infty}\frac{\text{d}t\,\text{d}\varepsilon }{2\pi}e^{i\varepsilon\hat{T}_{\mathcal{C}}}e^{-i\hat{H}_U t}\ket{\phi_{\mathcal{C}}(0)}\ket{\chi_{\mathcal{R}}}
   = \int_{-\infty}^{+\infty}\frac{\text{d}t\,\text{d}\varepsilon }{2\pi}e^{i\varepsilon\hat{T}_{\mathcal{C}}}e^{-i\hat{H}_U t}e^{-i\varepsilon\hat{T}_{\mathcal{C}}}e^{i\varepsilon\hat{T}_{\mathcal{C}}}\ket{\phi_{\mathcal{C}}(0)}\ket{\chi_{\mathcal{R}}}\\
    =\int_{-\infty}^{+\infty}\frac{\text{d}t\,\text{d}\varepsilon }{2\pi}e^{-it\left(\hat{H}_U +i\varepsilon\left[\hat{T}_{\mathcal{C}},\hat{H}_U\right]\right)}\ket{\phi_{\mathcal{C}}(0)}\ket{\chi_{\mathcal{R}}}
    =\int_{-\infty}^{+\infty}\frac{\text{d}t\,\text{d}\varepsilon }{2\pi}e^{-i\hat{H}_U t}e^{i\varepsilon t\hat{\alpha}}\ket{\phi_{\mathcal{C}}(0)}\ket{\chi_{\mathcal{R}}}
    =\int_{-\infty}^{+\infty}\text{d}t\,e^{-i\hat{H}_U t} \delta\left(t\hat{\alpha}\right)\ket{\phi_{\mathcal{C}}(0)}\ket{\chi_{\mathcal{R}}}\\
    =|\hat{\alpha}|^{+}\int_{-\infty}^{+\infty}\text{d}t\,e^{-i\hat{H}_U t} \delta\left(t\right)\ket{\phi_{\mathcal{C}}(0)}\ket{\chi_{\mathcal{R}}}
    =|\hat{\alpha}|^{+}\ket{\phi_{\mathcal{C}}(0)}\ket{\chi_{\mathcal{R}}},
\end{multline}
where I used Eqs.~\eqref{eq:commutator1} and \eqref{eq:cond1_NEW} in the third equality, Eq.~\eqref{eq:cond2_NEW} in the fourth one, and the fact that $\ket{\phi_{\mathcal{C}}(0)}\ket{\chi_{\mathcal{R}}}$ does not lie in the kernel of $\hat{\alpha}$ in the sixth one. This means that we can choose to construct $\kket{\Psi_U}$ starting from $|\hat{\alpha}|\ket{\phi_{\mathcal{C}}(0)}\ket{\chi_{\mathcal{R}}}$ (which is still of the form $\ket{\phi_{\mathcal{C}}(0)}\ket{\chi'_{\mathcal{R}}}$ because $\hat{\alpha}$ commutes with $\hat{T}_{\mathcal{C}}$, which is non-degenerate) so that $\hat{\Pi}_0\kket{\Psi_U}=\ket{\phi_{\mathcal{C}}(0)}\ket{\chi_{\mathcal{R}}}$ for any desired $\ket{\chi_{\mathcal{R}}}\in\mathscr{H}_{\mathcal{R}}$ such that $\ket{\phi_{\mathcal{C}}(0)}\ket{\chi_{\mathcal{R}}}$ does not lie in the kernel of $\hat{\alpha}$.

\section{Constraint equivalence}\label{appA:constraint_equivalence}

\paragraph{Hypotheses and weaker constraint.}
Consider some $\kket{\Psi_U}$ 0-eigenstate of $\hat{H}_U$ such that $\brakket{\phi_{\mathcal{C}}(t)}{\Psi_U}=\ket{\psi_{\mathcal{R}}(t)}$ evolves unitarily, that is, $\ket{\psi_{\mathcal{R}}(t)}=\hat{U}(t,t_0)\ket{\psi_{\mathcal{R}}(t_0)}$ for some $t_0\in \mathds{R}$, with $\hat{U}(t,t_0)$ a strongly continuous and differentiable unitary operator such that $\hat{U}(t,t_0)=\hat{U}(t,s)\hat{U}(s,t_0)$ $\forall s\in\mathds{R}$. This means that $i \frac{\text{d}\ket{\psi_{\mathcal{R}}(t)}}{\text{d}t} = \hat{X}_{\mathcal{R}}(t)\ket{\psi_{\mathcal{R}}(t)}$ $\forall t$, with $\hat{X}_{\mathcal{R}}(t)\coloneqq i \frac{\text{d}\hat{U}(t,t_0)}{\text{d}t}\hat{U}^{\dagger}(t,t_0)$ a self-adjoint operator. Therefore, $\kket{\Psi_U}$, which can also be written as $\kket{\Psi_U}=\int_{-\infty}^{+\infty}\text{d}t\,\ket{\phi_{\mathcal{C}}(t)}\ket{\psi_{\mathcal{R}}(t)}$, satisfies the constraint
\begin{equation}
    \left(\hat{H}_{\mathcal{C}}+\hat{X}_{\mathcal{R}}(\hat{T}_{\mathcal{C}})\right)\kket{\Psi_U}=0,
    \label{eq:constraint_new}
\end{equation}
with $\hat{X}_{\mathcal{R}}(\hat{T}_{\mathcal{C}})=\int_{-\infty}^{+\infty}\text{d}t\,\ket{\phi_{\mathcal{C}}(t)}\bra{\phi_{\mathcal{C}}(t)}\otimes \hat{X}_{\mathcal{R}}(t)$ \textit{symmetric}. Under the reasonable physical assumption $\sup\limits_{t}\norm{\hat{X}_{\mathcal{R}}(t)}<\infty$, $\hat{X}_{\mathcal{R}}(\hat{T}_{\mathcal{C}})$ is bounded and thus $\hat{H}_{\mathcal{C}}+\hat{X}_{\mathcal{R}}(\hat{T}_{\mathcal{C}})$ is self-adjoint on $\mathcal{D}(\hat{H}_{\mathcal{C}})$.
We can check explicitly that $\kket{\Psi_U}$ satisfies Eq.~\eqref{eq:constraint_new} in the following way
\begin{multline}
    \left(\hat{H}_{\mathcal{C}}+\hat{X}_{\mathcal{R}}(\hat{T}_{\mathcal{C}})\right)\kket{\Psi_U}
    =\int_{-\infty}^{+\infty}\text{d}t\,\left(\hat{H}_{\mathcal{C}}+\hat{X}_{\mathcal{R}}(\hat{T}_{\mathcal{C}})\right)\ket{\phi_{\mathcal{C}}(t)}\ket{\psi_{\mathcal{R}}(t)}\\
    =\int_{-\infty}^{+\infty}\text{d}t\left(i\frac{\text{d}\ket{\phi_{\mathcal{C}}(t)}}{\text{d}t}\ket{\psi_{\mathcal{R}}(t)} + \ket{\phi_{\mathcal{C}}(t)}\hat{X}_{\mathcal{R}}(t)\ket{\psi_{\mathcal{R}}(t)}\right)=\int_{-\infty}^{+\infty}\text{d}t\,\ket{\phi_{\mathcal{C}}(t)}\left(-i\frac{\text{d}}{\text{d}t}+\hat{X}_{\mathcal{R}}(t)\right)\ket{\psi_{\mathcal{R}}(t)}=0,
    \label{eq:eigenstate_C}
\end{multline}
where in the last equality I integrated by parts the first term and neglected the boundary terms, which are an artifact of working with the improper time states \cite{kuypers_measuring_2025}.
Therefore, if $\ket{\psi_{\mathcal{R}}(t)}=\brakket{\phi_{\mathcal{C}}(t)}{\Psi_U}$ evolves unitarily, then $\kket{\Psi_U}$ is also a 0-eigenstate of the constraint $\hat{\mathscr{C}}\coloneqq \hat{H}_{\mathcal{C}}+\hat{X}_{\mathcal{R}}(\hat{T}_{\mathcal{C}})$,  for which Eqs.~\eqref{eq:cond1_NEW}--\eqref{eq:cond2_NEW} hold. 

For the dynamics to be truly unitary, all the \textit{allowed} states $\ket{\psi'_{\mathcal{R}}(t)}=\brakket{\phi_{\mathcal{C}}(t)}{\Psi'_U}$ must evolve with the \textit{same} unitary $\hat{U}(t,t_0)$ and so all the stationary states $\kket{\Psi'_U}$ must also satisfy Eq.~\eqref{eq:constraint_new}. This means that the \textit{physical} Hilbert space $\mathscr{H}_{phy}(\hat{H}_U)$, i.e. the 0-eigenspace of $\hat{H}_U$, must be contained in $\mathscr{H}_{phy}(\hat{\mathscr{C}})$. Moreover, $\mathscr{H}_{phy}(\hat{H}_U)$ is an (improper) subspace of $\mathscr{H}_{phy}(\hat{\mathscr{C}})$ because it is closed under addition and multiplication by a scalar. 

\paragraph{Characterization of $\hat{H}_U$.}
Now, we can write  $\mathscr{H}_{phy}(\hat{\mathscr{C}})=\mathscr{H}_{phy}(\hat{H}_U)\oplus\left(\mathscr{H}^{\perp}_{phy}(\hat{H}_U)\cap\mathscr{H}_{phy}(\hat{\mathscr{C}})\right)$, with $\mathscr{H}^{\perp}_{phy}(\hat{H}_U)$ the orthogonal complement of $\mathscr{H}_{phy}(\hat{H}_U)$. We can easily characterize $\mathscr{H}^{\perp}_{phy}(\hat{H}_U)\cap\mathscr{H}_{phy}(\hat{\mathscr{C}})$ because a state $\kket{\Theta_U}\in\mathscr{H}_{phy}(\hat{\mathscr{C}}) $ also belongs to $\mathscr{H}^{\perp}_{phy}(\hat{H}_U)$ if $\bbrakket{\Theta_U}{\Psi_U}=0$ for any $\kket{\Psi_U}\in\mathscr{H}_{phy}(\hat{H}_U)$, but 
\begin{equation}
    \bbrakket{\Theta_U}{\Psi_U}=\int_{-\infty}^{+\infty}\text{d}t\,\braket{\theta_{\mathcal{R}}(t)}{\psi_{\mathcal{R}}(t)}
    =\left(\int_{-\infty}^{+\infty}\text{d}t\right)\braket{\theta_{\mathcal{R}}(t_0)}{\psi_{\mathcal{R}}(t_0)},
\end{equation} 
where the second equality follows from the fact that both $\ket{\theta_{\mathcal{R}}(t)}=\brakket{\phi_{\mathcal{C}}(t)}{\Theta_U}$ and $\ket{\psi_{\mathcal{R}}(t)}=\brakket{\phi_{\mathcal{C}}(t)}{\Psi_U}$ must evolve with the same unitary $\hat{U}(t,t_0)$. So $\bbrakket{\Theta_U}{\Psi_U}$ can be zero only if $\braket{\theta_{\mathcal{R}}(t_0)}{\psi_{\mathcal{R}}(t_0)}=0$ for some $t_0$. Therefore, we can associate $\mathscr{H}^{\perp}_{phy}(\hat{H}_U)\cap\mathscr{H}_{phy}(\hat{\mathscr{C}})$ with a projector $\hat{P}^{(0)}_{\mathcal{R}}(t_0)$ on the orthogonal complement of the subspace spanned by all the $\ket{\psi_{\mathcal{R}}(t_0)}$ compatible with $\mathscr{H}_{phy}(\hat{H}_U)$ at a certain time $t_0$. In other terms, the difference between $\mathscr{H}_{phy}(\hat{\mathscr{C}})$ and $\mathscr{H}_{phy}(\hat{H}_U)$ consists simply of a restriction on the allowed states $\ket{\psi_{\mathcal{R}}(t)}$ (while $\mathscr{H}_{phy}(\hat{\mathscr{C}})$ allows any state $\ket{\theta_{\mathcal{R}}(t)}\in\mathscr{H}_{\mathcal{R}}$ as discussed in Appendix \ref{appA:freedom}).

If we call $\hat{P}^{(+)}_{\mathcal{R}}(t_0)$ the projector on the orthogonal complement of $\hat{P}^{(0)}_{\mathcal{R}}(t_0)$, so that $\hat{P}^{(+)}_{\mathcal{R}}(t_0)+\hat{P}^{(0)}_{\mathcal{R}}(t_0)=\hat{\mathds{1}}_{\mathcal{R}}$ and $\hat{P}^{(+)}_{\mathcal{R}}(t_0)\hat{P}^{(0)}_{\mathcal{R}}(t_0)=\hat{P}^{(0)}_{\mathcal{R}}(t_0)\hat{P}^{(+)}_{\mathcal{R}}(t_0)=0$, the most general 0-eigenstate of $\hat{H}_U$ can be written as
\begin{multline}
    \kket{\Psi_U}=\int_{-\infty}^{+\infty}\text{d}t\,\ket{\phi_{\mathcal{C}}(t)}\hat{U}(t,t_0)\hat{P}^{(+)}_{\mathcal{R}}(t_0)\ket{\psi_{\mathcal{R}}(t_0)}
    =\int_{-\infty}^{+\infty}\text{d}t\,\ket{\phi_{\mathcal{C}}(t)}\hat{U}(t,t_0)\hat{P}^{(+)}_{\mathcal{R}}(t_0)\hat{U}^{\dagger}(t,t_0)\hat{U}(t,t_0)\ket{\psi_{\mathcal{R}}(t_0)}\\
    =\hat{P}^{(+)}(\hat{T}_{\mathcal{C}})\int_{-\infty}^{+\infty}\text{d}t\,\ket{\phi_{\mathcal{C}}(t)}\hat{U}(t,t_0)\ket{\psi_{\mathcal{R}}(t_0)}=\hat{P}^{(+)}(\hat{T}_{\mathcal{C}})\kket{\Psi_U},
\end{multline}
for any $\ket{\psi_{\mathcal{R}}(t_0)}\in\mathscr{H}_{\mathcal{R}}$ and $t_0\in\mathds{R}$, with $\hat{P}^{(+)}(\hat{T}_{\mathcal{C}})=\int_{-\infty}^{+\infty}\text{d}t\,\ket{\phi_{\mathcal{C}}(t)}\bra{\phi_{\mathcal{C}}(t)}\otimes \hat{P}^{(+)}_{\mathcal{R}}(t,t_0)$ where $\hat{P}^{(+)}_{\mathcal{R}}(t,t_0)= \hat{U}(t,t_0)\hat{P}^{(+)}_{\mathcal{R}}(t_0)\hat{U}^{\dagger}(t,t_0)$. 

\paragraph{Properties of $\hat{P}^{(+)}(\hat{T}_{\mathcal{C}})$.}
Since $\hat{U}(t,t_0)$ is a strongly continuous unitary and $\hat{P}^{(+)}_{\mathcal{R}}(t_0)$ is bounded (because $\mathcal{R}$ is finite-dimensional), $t\rightarrow\hat{P}^{(+)}_{\mathcal{R}}(t,t_0)$ is strongly continuous (and hence measurable), ensuring $\hat{P}^{(+)}(\hat{T}_{\mathcal{C}})$ is well-defined. Also note that $\hat{P}^{(+)}(\hat{T}_{\mathcal{C}})$ is a bounded, self-adjoint projector and it strongly commutes with $\hat{T}_{\mathcal{C}}$ because their spectral measures commute. Moreover, it is independent of $\hat{T}_{\mathcal{C}}$ if $\left[\hat{U}(t,t_0),\hat{P}^{(+)}_{\mathcal{R}}(t_0)\right]=0$ or, equivalently, $\left[\hat{X}_{\mathcal{R}}(t),\hat{P}^{(+)}_{\mathcal{R}}(t_0)\right]=0$, $\forall t \in \mathds{R}$. Finally, let us consider the commutator $\left[\hat{H}_{\mathcal{C}} +\hat{X}_{\mathcal{R}}(\hat{T}_{\mathcal{C}}), \hat{P}^{(+)}(\hat{T}_{\mathcal{C}})\right]$. First, we can calculate 
\begin{multline}
    \hat{H}_\mathcal{C}\hat{P}^{(+)}(\hat{T}_{\mathcal{C}})\kket{\Theta_U}=-i\int \text{d}t\,\ket{\phi_{\mathcal{C}}(t)}\left( \dot{\hat{P}}^{(+)}_{\mathcal{R}}(t,t_0)\ket{\theta_{\mathcal{R}}(t)}+\hat{P}^{(+)}_{\mathcal{R}}(t,t_0)\ket{\dot{\theta}_{\mathcal{R}}(t)}\right) \\
    =-i\int \text{d}t\,\ket{\phi_{\mathcal{C}}(t)}\left( -i\left[\hat{X}_{\mathcal{R}}(t),\hat{P}^{(+)}_{\mathcal{R}}(t,t_0)\right]\ket{\theta_{\mathcal{R}}(t)}+\hat{P}^{(+)}_{\mathcal{R}}(t,t_0)\ket{\dot{\theta}_{\mathcal{R}}(t)}\right)\\
    =-\left[\hat{X}(\hat{T}_{\mathcal{C}}),\hat{P}^{(+)}(\hat{T}_{\mathcal{C}})\right]\kket{\Theta_U} + \hat{P}^{(+)}\hat{H}_\mathcal{C}(\hat{T}_{\mathcal{C}})\kket{\Theta_U}
    \label{eq:HCP}
\end{multline}
where $\ket{\theta_{\mathcal{R}}(t)}\coloneqq \brakket{\phi_{\mathcal{C}}(t)}{\Theta_U}$. From this we can deduce that 
\begin{multline}
    \norm{\hat{H}_\mathcal{C}\hat{P}^{(+)}(\hat{T}_{\mathcal{C}})\kket{\Theta_U}}\leq \norm{\left[\hat{X}(\hat{T}_{\mathcal{C}}),\hat{P}^{(+)}(\hat{T}_{\mathcal{C}})\right]\kket{\Theta_U}}+\norm{\hat{P}^{(+)}\hat{H}_\mathcal{C}\kket{\Theta_U}}\\
    \leq \norm{\hat{X}(\hat{T}_{\mathcal{C}})} \cdot \norm{\hat{P}^{(+)}(\hat{T}_{\mathcal{C}})\kket{\Theta_U}}+ \norm{\hat{P}^{(+)}(\hat{T}_{\mathcal{C}})} \cdot \norm{\hat{X}(\hat{T}_{\mathcal{C}})\kket{\Theta_U}}+\norm{\hat{P}^{(+)}} \cdot \norm{\hat{H}_\mathcal{C}\kket{\Theta_U}}\\
    \leq 2\norm{\hat{X}(\hat{T}_{\mathcal{C}})} \cdot \norm{\vphantom{\hat{X}(\hat{T}_{\mathcal{C}})}\kket{\Theta_U}}+\norm{\hat{H}_\mathcal{C}\kket{\Theta_U}},
    \label{eq:norm_HCP}
\end{multline}
which is finite if $\kket{\Theta_U}\in \mathcal{D}(\hat{H}_{\mathcal{C}})$ and, as assumed before, $\sup\limits_{t}\norm{\hat{X}_{\mathcal{R}}(t)}<\infty$. This means that $\hat{P}^{(+)}(\hat{T}_{\mathcal{C}})\mathcal{D}(\hat{H}_\mathcal{C})\subseteq \mathcal{D}(\hat{H}_\mathcal{C})$ and therefore the commutator $ \left[\hat{H}_{\mathcal{C}} +\hat{X}_{\mathcal{R}}(\hat{T}_{\mathcal{C}}), \hat{P}^{(+)}(\hat{T}_{\mathcal{C}})\right]$ is well-defined on $\mathcal{D}(\hat{H}_\mathcal{C})$. 
Finally, from Eq.~\eqref{eq:HCP} we have that
\begin{equation}
    \left[\hat{H}_{\mathcal{C}} +\hat{X}_{\mathcal{R}}(\hat{T}_{\mathcal{C}}), \hat{P}^{(+)}(\hat{T}_{\mathcal{C}})\right]=0,
    \label{eq:comm_CP}
\end{equation}
on $\mathcal{D}(\hat{H}_\mathcal{C})$.

\paragraph{Constraint equivalence.}
We can use $\hat{P}^{(+)}(\hat{T}_{\mathcal{C}})$ to find a constraint that restricts $\mathscr{H}_{phy}(\hat{\mathscr{C}})$ and makes it coincide with $\mathscr{H}_{phy}(\hat{H}_U)$:
\begin{equation}
    \hat{\mathscr{C}}'\coloneqq \hat{P}^{(+)}(\hat{T}_{\mathcal{C}})\left( \hat{H}_{\mathcal{C}} +\hat{X}_{\mathcal{R}}(\hat{T}_{\mathcal{C}}) \right) \hat{P}^{(+)}(\hat{T}_{\mathcal{C}}) + \hat{P}^{(0)}(\hat{T}_{\mathcal{C}}).
\end{equation}
$\hat{\mathscr{C}}'$ is block-diagonal with respect to the decomposition induced by $\hat{P}^{(+)}(\hat{T}_{\mathcal{C}})$ and is self-adjoint on $\mathcal{D}(\hat{\mathscr{C}}')=\left\{ \kket{\Psi} : \hat{P}^{(+)}(\hat{T}_{\mathcal{C}})\kket{\Psi} \in \mathcal{D}(\hat{H}_{\mathcal{C}}) \right\}=\hat{P}^{(+)}(\hat{T}_{\mathcal{C}})\mathcal{D}(\hat{H}_{\mathcal{C}})\oplus \text{Ran}\left(\hat{P}^{(0)}(\hat{T}_{\mathcal{C}})\right)$. Moreover, $\hat{P}^{(+)}(\hat{T}_{\mathcal{C}})$ is a reducing projection for $\hat{\mathscr{C}}'$, meaning that $\hat{P}^{(+)}(\hat{T}_{\mathcal{C}})\mathcal{D}(\hat{\mathscr{C}}')\subseteq \mathcal{D}(\hat{\mathscr{C}}')$ and $\hat{\mathscr{C}}'\hat{P}^{(+)}(\hat{T}_{\mathcal{C}})\kket{\Theta}=\hat{P}^{(+)}(\hat{T}_{\mathcal{C}})\hat{\mathscr{C}}'\kket{\Theta}$, $\forall \kket{\Theta}\in \mathcal{D}(\hat{\mathscr{C}}')$. Therefore, $\hat{P}^{(+)}(\hat{T}_{\mathcal{C}})$ and $\hat{\mathscr{C}}'$ strongly commute.
The rate operator associated with this constraint is
\begin{equation}
    \hat{\alpha}=i \left[\hat{\mathscr{C}}',\hat{T}_{\mathcal{C}}\right]=\hat{P}^{(+)}(\hat{T}_{\mathcal{C}}),
\end{equation}
which, as discussed above, strongly commutes with both $\hat{T}_{\mathcal{C}}$ and $\hat{\mathscr{C}}'$.

Now we can check that any 0-eigenstate $\kket{\Psi_U}=\hat{P}^{(+)}(\hat{T}_{\mathcal{C}})\kket{\Psi_U}$ of $\hat{H}_U$ is also a 0-eigenstate of $\hat{\mathscr{C}}'$:
\begin{equation}
    \hat{\mathscr{C}}'\kket{\Psi_U}= \hat{P}^{(+)}(\hat{T}_{\mathcal{C}})\left( \hat{H}_{\mathcal{C}} +\hat{X}_{\mathcal{R}}(\hat{T}_{\mathcal{C}}) \right)\kket{\Psi_U} =0,
\end{equation}
where I have used Eq.~\eqref{eq:constraint_new}.
Therefore, we have that $\mathscr{H}_{phy}(\hat{H}_U)\subseteq\mathscr{H}_{phy}(\hat{\mathscr{C}}')$.
Conversely, if $\kket{\Phi_U}$ is a 0-eigenstate of $\hat{\mathcal{C}}'$ then we can multiply $\hat{\mathcal{C}}'\kket{\Phi_U}=0$ by $\hat{P}^{(+)}(\hat{T}_{\mathcal{C}})$ and $\hat{P}^{(0)}(\hat{T}_{\mathcal{C}})$ on the left to get
\begin{gather}
    \hat{P}^{(+)}(\hat{T}_{\mathcal{C}})\left( \hat{H}_{\mathcal{C}} +\hat{X}_{\mathcal{R}}(\hat{T}_{\mathcal{C}}) \right) \hat{P}^{(+)}(\hat{T}_{\mathcal{C}})\kket{\Phi_U} =0, \label{eq:P+C'}\\
     \hat{P}^{(0)}(\hat{T}_{\mathcal{C}})\kket{\Phi_U}=0. \label{eq:P0C'}
\end{gather}
The second equation implies that  $\kket{\Phi_U}=\hat{P}^{(+)}(\hat{T}_{\mathcal{C}})\kket{\Phi_U}$ and so the first equation, using Eq.~\eqref{eq:comm_CP},  becomes
\begin{equation}
    \left( \hat{H}_{\mathcal{C}} +\hat{X}_{\mathcal{R}}(\hat{T}_{\mathcal{C}}) \right) \hat{P}^{(+)}(\hat{T}_{\mathcal{C}})\kket{\Phi_U} =\left( \hat{H}_{\mathcal{C}} +\hat{X}_{\mathcal{R}}(\hat{T}_{\mathcal{C}}) \right)\kket{\Phi_U}=0.
    \label{eq:P+C'_new}
\end{equation}
Taken together, Eqs.~\eqref{eq:P+C'_new} and \eqref{eq:P0C'} are the same equations that characterise the 0-eigenstates of $\hat{H}_U$. Therefore, we also have that $\mathscr{H}_{phy}(\hat{\mathscr{C}}')\subseteq\mathscr{H}_{phy}(\hat{H}_U)$ and so  $\mathscr{H}_{phy}(\hat{\mathscr{C}}')=\mathscr{H}_{phy}(\hat{H}_U)$. We have thus proved that if $\ket{\psi_{\mathcal{R}}(t)}$ evolves unitarily then $\hat{H}_U$ must be physically equivalent to a constraint for which Eqs.~\eqref{eq:cond1_NEW}-\eqref{eq:cond2_NEW} hold.

\section{Rate and variance of the clock}\label{appA:clock}
\paragraph{Rate of the clock.}
Given the Hamiltonian of the Universe  $\hat{H}_U=\hat{H}_{\mathcal{C}} + \hat{H}_{\mathcal{R}}+\hat{V}_{\mathcal{C}\mathcal{R}} + \hat{H}_{\mathcal{C}_2}$, and the state of $\mathcal{CR}$ when $\mathcal{C}_2$ shows the time $t$, $\ket{\psi_{\mathcal{CR}}(t)}=e^{-i\hat{H}_{\mathcal{CR}}t}\ket{\psi_{\mathcal{CR}}(0)}$, with $\hat{H}_{\mathcal{CR}}\coloneqq \hat{H}_{\mathcal{C}} + \hat{H}_{\mathcal{R}}+\hat{V}_{\mathcal{C}\mathcal{R}}$,
the rate of $\mathcal{C}$ from the perspective of $\mathcal{C}_2$ is
\begin{equation}
\alpha(t)=\frac{\text{d}\overline{\tau_{\mathcal{C}}}(t)}{\text{d}t}=\frac{\text{d}}{\text{d}t}\frac{\bra{\psi_{\mathcal{CR}}(t)}\hat{T}_{\mathcal{C}}\ket{\psi_{\mathcal{CR}}(t)}}{\braket{\psi_{\mathcal{CR}}(t)}{\psi_{\mathcal{CR}}(t)}}= \frac{\bra{\psi_{\mathcal{CR}}(t)}i\left[\hat{H}_{\mathcal{CR}}, \hat{T}_{\mathcal{C}}\right] \ket{\psi_{\mathcal{CR}}(t)}}{\braket{\psi_{\mathcal{CR}}(0)}{\psi_{\mathcal{CR}}(0)} } 
=\frac{\bra{\psi_{\mathcal{CR}}(t)} \hat{\alpha} \ket{\psi_{\mathcal{CR}}(t)}}{\braket{\psi_{\mathcal{CR}}(0)}{\psi_{\mathcal{CR}}(0)}},
    \label{eq:alpha_t}
\end{equation}
where I used the fact that $ \ket{\psi_{\mathcal{CR}}(t)}$ evolves unitarily so $\braket{\psi_{\mathcal{CR}}(t)}{\psi_{\mathcal{CR}}(t)}=\braket{\psi_{\mathcal{CR}}(0)}{\psi_{\mathcal{CR}}(0)}$.
Taking the derivative of $\alpha(t)$ with respect to $t$ leads to
\begin{equation}
\frac{\text{d}\alpha(t)}{\text{d}t}=\frac{\bra{\psi_{\mathcal{CR}}(t)}  i\left[\hat{H}_{\mathcal{CR}}, \hat{\alpha}\right]\ket{\psi_{\mathcal{CR}}(t)}}{\braket{\psi_{\mathcal{CR}}(0)}{\psi_{\mathcal{CR}}(0)} }=0,    
\end{equation}
where the last equation follows from Eq.~\eqref{eq:cond2_NEW}.

\paragraph{Variance of the clock readings.}
The variance of the clock readings from the perspective of $\mathcal{C}_2$ is given by
\begin{equation}
     \sigma^2_{\tau_{\mathcal{C}}}(t)\coloneqq\frac{\bra{\psi_{\mathcal{CR}}(t)}\hat{T}_{\mathcal{C}}^2\ket{\psi_{\mathcal{CR}}(t)}}{\braket{\psi_{\mathcal{CR}}(t)}{\psi_{\mathcal{CR}}(t)}}-\overline{\tau_{\mathcal{C}}}(t)^2.
     \label{eq:variance}
\end{equation}
The first term can be calculated using the expansion
\begin{equation}
    e^{i\hat{H}_{\mathcal{CR}}t}\hat{T}^2_{\mathcal{C}}e^{- i\hat{H}_{\mathcal{CR}}t}=\hat{T}^2_{\mathcal{C}}+it\left[\hat{H}_{\mathcal{CR}},\hat{T}^2_{\mathcal{C}}\right]
    +\frac{(it)^2}{2!} \left[\hat{H}_{\mathcal{CR}},\left[\hat{H}_{\mathcal{CR}},\hat{T}^2_{\mathcal{C}}\right]\right] + \dots,
    \label{eq:expansion_T2}
\end{equation}
and 
\begin{gather}
    \left[\hat{H}_{\mathcal{CR}},\hat{T}^2_{\mathcal{C}}\right]=\left[\hat{H}_{\mathcal{CR}},\hat{T}_{\mathcal{C}}\right]\hat{T}_{\mathcal{C}}+\hat{T}_{\mathcal{C}}\left[\hat{H}_{\mathcal{CR}},\hat{T}_{\mathcal{C}}\right]=-i\hat{\alpha}\hat{T}_{\mathcal{C}}-i\hat{T}_{\mathcal{C}}\hat{\alpha},\\ 
    \left[\hat{H}_{\mathcal{CR}},\left[\hat{H}_{\mathcal{CR}},\hat{T}^2_{\mathcal{C}}\right]\right]=\left[\hat{H}_{\mathcal{CR}},-i\hat{\alpha}\hat{T}_{\mathcal{C}}-i\hat{T}_{\mathcal{C}}\hat{\alpha}\right]\
    =-i\hat{\alpha}\left[\hat{H}_{\mathcal{CR}},\hat{T}_{\mathcal{C}}\right]-i\left[\hat{H}_{\mathcal{CR}},\hat{T}_{\mathcal{C}}\right]\hat{\alpha}=-2\hat{\alpha}^2,
    \label{eq:expansion_T2_ord2}
\end{gather}
where \eqref{eq:expansion_T2_ord2} uses Eq.~\eqref{eq:cond2_NEW}. Eqs.~\eqref{eq:cond2_NEW} and \eqref{eq:expansion_T2_ord2} also imply that the terms of order $t^3$ and above in Eq.~\eqref{eq:expansion_T2} are zero.
Putting these results in Eqs.~\eqref{eq:variance} and using Eqs.~\eqref{eq:time_transformation} we get
\begin{equation}
      \sigma^2_{\tau_{\mathcal{C}}}(t)
      \coloneqq \expval{\hat{T}_{\mathcal{C}}^2 +t(\hat{\alpha}\hat{T}_{\mathcal{C}}+\hat{T}_{\mathcal{C}}\hat{\alpha}) + t^2 \hat{\alpha}^2} - \left(\expval{\hat{T}_{\mathcal{C}}}+\expval{\hat{\alpha}}t \right)^2
      =\sigma^2_{\tau_{\mathcal{C}}}(0)+t^2 \sigma^2_{\alpha}+2t\,\text{Cov}(\alpha(0),\tau_{\mathcal{C}}(0)),
\end{equation}
with $\expval{\cdot}=\frac{\bra{\psi_{\mathcal{CR}}(0)}\cdot\ket{\psi_{\mathcal{CR}}(0)}}{\braket{\psi_{\mathcal{CR}}(0)}{\psi_{\mathcal{CR}}(0)}}$,
$\sigma^2_{\tau_\mathcal{C}}\coloneqq \expval{\hat{T}_{\mathcal{C}}^2}-\expval{\hat{T}_{\mathcal{C}}}^2$, $\sigma^2_{\alpha}\coloneqq \expval{\hat{\alpha}^2}-\expval{\hat{\alpha}}^2$ and 
$\text{Cov}(\alpha(0),\tau_{\mathcal{C}}(0))\coloneqq \frac{1}{2}\expval{\hat{\alpha}\hat{T}_{\mathcal{C}}}+ \frac{1}{2}\expval{\hat{T}_{\mathcal{C}}\hat{\alpha}}-\expval{\hat{\alpha}}\expval{\hat{T}_{\mathcal{C}}}$.
When $\ket{\psi_{\mathcal{CR}}(0)}$ corresponds to a definite reading of the clock, i.e., it is an eigenstate of $\hat{T}_{\mathcal{C}}$, we have 
\begin{equation}
    \sigma^2_{\tau_{\mathcal{C}}}(t)=t^2 \sigma^2_{\alpha}.
\end{equation}


\end{document}